\documentclass[twocolumn,prb,reprint,superscriptaddress,floatfix,amsmath,amssymb,aps]{revtex4}

\usepackage[utf8]{inputenc}
\usepackage{graphicx}
\usepackage{float}
\usepackage{amsmath}
\usepackage{dcolumn}
\usepackage{bm}
\usepackage{xcolor}
\usepackage{xfrac}
\usepackage[colorlinks=true, allcolors=blue]{hyperref}
\usepackage{chemformula}
\usepackage{siunitx}

\DeclareSIUnit\oersted{Oe}
\DeclareSIUnit\emu{emu}
\DeclareSIUnit\erg{erg}
\DeclareSIUnit\gauss{G}
\DeclareSIUnit\bohr{\mu_B}
\DeclareSIUnit\atom{atom}
\DeclareSIUnit\molatom{mol_{atom}}
\def\muSR{$\mu$SR}

\begin{document}

\title{Magnetic properties of equiatomic CrMnFeCoNi}

\author{Timothy A.\ Elmslie}
\affiliation{Department of Physics, University of Florida, Gainesville, FL 32611, USA}
\author{Jacob Startt}
\affiliation{Center for Integrated Nanotechnologies, Sandia National Laboratories, Albuquerque, NM, 87185, USA}
\author{Sujeily Soto-Medina}
\affiliation{Department of Materials Science and Engineering, University of Florida, Gainesville, FL 32611, USA}
\author{Yang Yang}
\affiliation{Department of Materials Science and Engineering, University of Florida, Gainesville, FL 32611, USA}
\author{Keke Feng}
\affiliation{Department of Physics, Florida State University and National High Magnetic Field Laboratory, Tallahassee Florida 32310, USA}
\author{Ryan E.\ Baumbach}
\affiliation{Department of Physics, Florida State University and National High Magnetic Field Laboratory, Tallahassee Florida 32310, USA}
\author{Emma Zappala}
\affiliation{Department of Physics and Astronomy, Brigham Young University, Provo, Utah 84602, USA}
\author{Gerald D.\ Morris}
\affiliation{Centre for Molecular and Materials Science, TRIUMF, Vancouver, British Columbia V6T 2A3, Canada}
\author{Benjamin A.\ Frandsen}
\affiliation{Department of Physics and Astronomy, Brigham Young University, Provo, Utah 84602, USA.} 
\author{Mark W.\ Meisel}
\affiliation{Department of Physics and the National High Magnetic Field Laboratory, University of Florida, Gainesville, Florida 32611, USA}
\author{Michele V.\ Manuel}
\affiliation{Department of Materials Science and Engineering, University of Florida, Gainesville, FL 32611, USA}
\author{R\'emi Dingreville}
\affiliation{Center for Integrated Nanotechnologies, Sandia National Laboratories, Albuquerque, NM, 87185, USA}
\author{James J.\ Hamlin}
\affiliation{Department of Physics, University of Florida, Gainesville, FL 32611, USA}

\date{\today}

\begin{abstract}
    Magnetic, specific heat, and structural properties of the equiatomic Cantor alloy system are reported for temperatures between \SI{5}{\kelvin} and \SI{300}{\kelvin}, and up to fields of \SI{70}{\kilo \oersted}.
    Magnetization measurements performed on as-cast, annealed, and cold-worked samples reveal a strong processing history dependence and that high-temperature annealing after cold-working does not restore the alloy to a ``pristine'' state.
    Measurements on known precipitates show that the two transitions, detected at \SI{43}{\kelvin} and \SI{85}{\kelvin}, are intrinsic to the Cantor alloy and not the result of an impurity phase.
    Experimental and \emph{ab initio} density functional theory (DFT) computational results suggest that these transitions are a weak ferrimagnetic transition and a spin-glass-like transition, respectively, and magnetic and specific heat measurements provide evidence of significant Stoner enhancement and electron-electron interactions within the material.
\end{abstract}

\maketitle

\section{Introduction} \label{intro}
Investigation of high-entropy alloys began in the early 2000s with the work of Cantor, Kim, and Warren~\cite{cantor_novel_2002}, who investigated multicomponent alloys of the formula $A_xB_{90-x}\mathrm{Al}_{10}$, in which $A$ and $B$ each represented a combination of two to four different elements in equiatomic ratios.
In this way, they were able to produce stable combinations of up to seven different elements.
The phrase ``high-entropy alloy,'' however, was not introduced until 2004, in the work of Yeh \emph{et al.}~\cite{yeh_nanostructured_2004}, who used it to describe compounds consisting of five or more elements.
These alloys are of interest partly due to the extremely large number of new alloy systems which fall into the category of high or medium entropy alloys~\cite{yeh_nanostructured_2004}.
Furthermore, prior investigations have revealed a number of alloys with intriguing properties such as high hardness and resistance to anneal softening~\cite{yeh_nanostructured_2004}, shape memory effects~\cite{firstov_high_2015,chen_shape_2019}, and superconductivity~\cite{motla_boron_2022}.
Cantor alloys also have potential applications due to soft magnetism~\cite{zhang_high-entropy_2013} and other tunable magnetic properties~\cite{zuo_tailoring_2017}.

The specific combination of chromium (Cr), manganese (Mn), iron (Fe), cobalt (Co), and nickel (Ni) that would come to characterize the ``Cantor alloy'' was first identified in the 2004 work of Cantor, Chang, Knight, and Vincent~\cite{cantor_microstructural_2004}, who produced alloys of sixteen and twenty different elements in equal proportions through induction melting.
The resulting compound was unsurprisingly multiphase, but the predominant phase was especially rich in Cr, Mn, Fe, Ni, and Co.
In the same work, this \ch{CrMnFeCoNi} compound was identified as face-centered cubic (FCC), and the ability of this material to dissolve large amounts of elements such as niobium (Nb), titanium (Ti), and vanadium (V) was noted.

Much work has been done regarding the mechanical properties of Cantor alloys~\cite{cantor_microstructural_2004,otto_influences_2013,laurent-brocq_insights_2015,jang_high-temperature_2018,kim_mechanical_2018,kim_-situ_2018}, as well as the magnetic properties of similar compounds~\cite{wang_novel_2007,zhang_annealing_2010,kao_electrical_2011,lucas_magnetic_2011,singh_effect_2011,liu_microstructure_2012,ma_effect_2012,zhang_effects_2012,tsai_physical_2013,zhang_high-entropy_2013,leong_effect_2017,zuo_tailoring_2017}, but fewer studies have examined the magnetic properties of the original \ch{CrMnFeCoNi} alloy~\cite{jin_tailoring_2016,yu_high-entropy_2016,schneeweiss_magnetic_2017,kamarad_effect_2019,billington_bulk_2020}.
Of its component elements, Cr and Mn are antiferromagnetic with $T_N = \SI{311}{\kelvin}$~\cite{bacon_magnetic_1969} and $T_N = \SI{100}{\kelvin}$~\cite{patrick_antiferromagnetism_1954} respectively, while Fe, Co, and Ni are ferromagnetic with $T_C = \SI{1043}{\kelvin}$, $T_C = \SI{1394}{\kelvin}$, and $T_C = \SI{631}{\kelvin}$, respectively~\cite{rumble_crc_2021}.
However, the magnetic transitions in the equiatomic Cantor alloy appear at much lower temperature.
Jin \emph{et al.}~\cite{jin_tailoring_2016} identified a peak in the magnetization of \ch{CrMnFeCoNi} at  \SI{25}{\kelvin} and suggested that it could be either an antiferromagnetic transition or a spin-glass transition.
Schneeweiss \emph{et al.}~\cite{schneeweiss_magnetic_2017}, however, found two transitions in the magnetization:  a spin-glass transition at \SI{93}{\kelvin}, and a ferromagnetic transition at \SI{38}{\kelvin}.
They were able to identify the \SI{38}{\kelvin} transition as ferromagnetic by magnetic hysteresis loops after cooling in field, while the nature of the spin-glass transition was confirmed by magnetic moment relaxation measurements.
Despite identifying these magnetic characteristics, the data of Schneeweiss \emph{et al.}\ do not afford additional quantitative analysis since the magnetization results are only reported in units of \si{\emu}, rather than in a generic manner like \si{\emu \per \gram}.
Consequently, the identification of the atoms and/or atomic morphology generating the magnetic response is not quantitatively addressed, and this work provides insights about this important issue.

Kamar\'ad \emph{et al.}~\cite{kamarad_effect_2019} investigated the magnetic properties of the equiatomic Cantor alloy between ambient pressure and \SI{1}{\giga \pascal}, observing that increasing pressure decreased magnetization slightly.
Using Curie-Weiss fitting, they found evidence of strong antiferromagnetic interactions, which were conjectured to be responsible for the observed small magnetization values as well as the linear field dependence of the magnetization.
In contrast, while Schneeweiss \emph{et al.}\ found ferromagnetic-type behavior in magnetic hysteresis loops after cooling in applied magnetic field, Kamar\'ad \emph{et al.}\ were unable to replicate these results, seeing a change of only \SI{0.4} {\emu \per \gram} in magnetization after field-cooling.
These observations led Kamar\'ad \emph{et al.}\ to disagree with Schneeweiss \emph{et al.}\ on the nature of the magnetic transitions within the compound, identifying ferrimagnetic order below \SI{85}{\kelvin} and magnetic cluster-glass behavior below \SI{43.5}{\kelvin}.

This work reports a quantitative analysis of the magnetic properties of the equiatomic Cantor alloy CrMnFeCoNi based on a combination of compositional and structural characterization, magnetization studies, Hall effect measurements, muon spin relaxation (\muSR) results, specific heat studies, and \emph{ab initio} density functional theory (DFT) calculations. 
The magnetic response of as-cast, annealed, and cold-worked samples revealed a strong history dependence, which is described and contrasted with the magnetic properties of precipitates known to form in Cantor alloys.
All of our results lead to the identification of the magnetic signatures as intrinsic characteristics of the equiatomic Cantor alloy CrMnFeCoNi, with weak ferrimagnetic order near \SI{43}{\kelvin}, a spin-glass-like fingerprint near \SI{85}{\kelvin}, and a sizable temperature-independent response.  
This latter contribution was established by Curie-Weiss analysis of the magnetic susceptibility data and further probed by Hall measurements to determine the expected size of the Pauli paramagnetic contribution to the susceptibility.  
Combined with an interpretation of the specific heat data and DFT results, our work provides evidence of a sizable Stoner enhancement and electron-electron interactions.

\section{Methods} \label{method}

\subsection{Materials Synthesis and History}
Samples were synthesized by combining stoichiometric amounts of elemental \ch{Cr}, \ch{Mn}, \ch{Fe}, \ch{Co}, and \ch{Ni} and melting them together in an Edmund B{\"u}hler MAM-1 compact arc melter to produce as-cast samples.
The Cr, Mn, Fe, and Ni elements were sourced from Alfa Aesar, while Co was purchased from Cerac.
The Cr used for synthesis was 99.995\% pure, while the Co was 99.5\% pure.
All other elements were 99.95\% pure.
Each sample was melted five times, flipping it over between each melt to improve sample homogeneity.
Samples measured immediately after arc-melting are referred to as ``as-cast.''
Annealed samples were made by sealing as-cast samples in quartz tubes under Ar atmosphere and homogenizing them at \SI{1100}{\celsius} for 6 days, after which the tube containing the samples was quenched in water.
Samples measured following this step are called ``anneal A.''
The piece cut from the as-cast boule and annealed in ambient atmosphere at \SI{700}{\celsius} for one hour is referred to as ``oxidized.''
Some samples from the anneal A batch were cold-worked by flattening them in a hydraulic press a total of three times using a pressure of approximately \SI{0.5}{\giga \pascal}, folding them in half between each flattening step.
These samples are known as ``cold-worked.''
After cold-working, the samples were re-annealed in quartz tubes under argon (Ar) atmosphere.
A portion were annealed at \SI{700}{\celsius} for one hour, while others were again subjected to \SI{1100}{\celsius} for six days, referred to as ``anneal B'' and ``anneal C,'' respectively.

\subsection{Compositional and Structural Analysis} \label{method:Comp}
The microstructure of the sample was characterized using a Tescan MIRA3 scanning electron microscope (SEM) with an energy-dispersive X-ray (EDX) detector at \SI{20}{\kilo\volt}.
Prior to SEM measurements, samples were mounted in resin epoxy and first polished using 600 grit, then 800 grit SiC paper.
This surface was further polished using alcohol-based lubricant and diamond paste of varying particulate sizes in steps of \SI{6}{\micro \meter}, \SI{3}{\micro \meter}, \SI{1}{\micro \meter} followed by \SI{0.05}{\micro \meter} master polishing using water-free colloidal silica suspension.
The crystal structure of the sample was investigated using a Panalytical Xpert Powder X-ray diffractometer (XRD) with a copper-radiation source at an accelerating voltage of \SI{40}{\kilo\volt} and an electron current of \SI{40}{\milli\ampere}, over a 2$\theta$ range from \ang{40} to \ang{120}.
Powder samples were prepared using a mortar and pestle.  
Quantitative composition data were obtained by electron probe microanalysis (EPMA) using a CAMECA SXFiveFE instrument operating with an accelerating voltage of \SI{15}{\kilo \volt} and a probe current of \SI{20}{\nano \ampere}.

\subsection{Magnetic and Magnetotransport Studies} \label{method:Mag}
A Quantum Design Magnetic Property Measurement System (MPMS) was used to take magnetization data.
Small pieces ranging from a few milligrams to a few hundred milligrams were cut from larger samples using an Allied 3000 low speed saw to minimize unintentional working of the samples.
Each sample was secured in a gel capsule inside a plastic straw for measurement.
In order to obtain magnetization versus temperature data, samples were cooled under zero field to a temperature of \SI{5}{\kelvin}, after which field was applied and data was recorded while warming.
This process produced the datasets labeled zero-field cooled (ZFC).
The applied field was held constant while the temperature warms to \SI{300}{\kelvin} and subsequently cools to \SI{5}{\kelvin} again.
Then data was recorded while warming to produce field-cooled warming (FCW) datasets.

Resistivity and Hall measurements were performed in a Quantum Design Physical Property Measurement System (PPMS) at \SI{300}{\kelvin}.
For these measurements, an Allied 3000 low speed saw was used to cut an \SI{8}{\milli \meter} by \SI{8.4}{\milli \meter} flat square sample from the arc-melted boule.
Then four notches, one on each side, were cut into it with the same device to produce a rough clover leaf shape for the purposes of Van der Pauw measurements.
To increase the signal-to-noise ratio, the sample was first sanded on a custom jig to a thickness of \SI{0.2}{\milli \meter}, then polished with 800 grit silicon carbide paper to a final thickness of \SI{0.16}{\milli \meter}.
This sample was placed on a PPMS transport puck with a piece of cigarette paper between sample and puck for insulation, while N-grease provided thermal contact.
Each ``leaf'' of the clover-like shape of the sample was connected to one of the puck solder pads using platinum wire affixed to the sample with silver paint.

\subsection{Muon spin relaxation experiments} \label{method:Muon}
Muon spin spectroscopy measurements, or \muSR, were conducted to probe the spin dynamics and provide sensitivity to the volume fraction of the various magnetic phases in equiatomic Cantor alloy.
These experiments were conducted at TRIUMF Laboratory in Vancouver, Canada using the LAMPF spectrometer on the M20D beamline. 
Positive muons with 100\% initial spin polarization were implanted one at a time in the sample.
After implantation, the muon spin underwent Larmor precession in the local magnetic field at the muon site, which consists of the vector sum of the internal magnetic field due to electronic and/or nuclear dipolar moments and any externally applied magnetic field. 
After a mean lifetime of 2.2~$\mu$s, the muon decays into two neutrinos and a positron, with the latter being emitted preferentially in the direction of the muon spin at the moment of decay. 
Two detectors placed on opposite sides of the sample record positron events as a function of time after muon implantation, yielding the \muSR\ asymmetry $a(t) = [N_1(t)-N_2(t)]/[N_1(t)+N_2(t)]$, where $N_1(t)$ and $N_2(t)$ are the number of positron events recorded at time $t$. 
This quantity is proportional to the projection of the muon ensemble spin polarization along the axis defined by the positions of the detectors, from which information about the local magnetic field distribution in the sample can be inferred~\cite{blund;cp10}. 
The sample was mounted on a low background copper sample holder, and the temperature controlled using a helium gas flow cryostat. 
The \muSR\ data were analyzed using the open source programs MusrFit~\cite{suter;physproc12} and BEAMS~\cite{peter;gh21}, which yield consistent results. 
The correction parameter $\alpha$ was determined according to standard practice from a weak transverse field measurement in the paramagnetic state~\cite{yaoua;b;msr11}.

\subsection{Specific Heat Measurements} \label{method:HC}
Alloy specific heat was measured in a PPMS with the attached PPMS heat capacity option in order to obtain the Debye temperature and the approximate magnetic entropy of the sample.
The sample used for specific heat measurements was annealed at \SI{1100}{\celsius} for six days before being cut to a mass of a few milligrams using an Allied 3000 low speed saw to minimize sample deformation and strain.
After cutting, this piece was placed on the sample platform of a Quantum Design heat capacity puck using a small amount of Apeizon N grease to adhere the sample and provide thermal contact before being inserted into the PPMS for measurement.

\subsection{DFT Calculations} \label{method:DFT}
Spin collinear DFT simulations were performed to investigate exchange splitting between spin states and its effect on ferromagnetism within the system.
For modeling, the Vienna ab-initio Simulation Package (VASP)~\cite{Kresse1993, Kresse1994, Kresse1995} was used, in which the electronic wavefunctions were modeled through plane-waves and projector-augmented wave (PAW) pseudopotentials~\cite{Blochl1994, Kresse1999}.
Exchange-correlation effects were treated within the generalized gradient approximation (GGA) according to the parametrization of Perdew, Burke, and Ernzerhof (PBE)~\cite{Perdew1996}, while the individual pseudopotentials used to model each species were chosen so that only the outermost $s$- and $d$-valence electrons were explicitly included.

The alloy atomic structure was modeled using supercells containing 108 atoms -- representing a $3\times3\times3$ transformation of the conventional four atom FCC unit cell. 
Special quasi-random structures (SQS)~\cite{Wei1990, Zunger1990}, built using the openly available Alloy-Theoretic Automated Toolkit (ATAT)~\cite{Vandewall2013}, were used to replicate random atomic ordering among lattice sites.
To ensure the most accurate description of properties of the randomly ordered solid solution structure, four separate 108 atom SQS supercells were modeled at the equiatomic composition, each using a $5\times5\times5$ gamma-entered k-point mesh.
For all simulations, electronic convergence was met when the total energy of the system fell below \SI{1.0E-6}{\eV}, while ionic convergence was met when forces on all atoms fall below \SI{20E-3}{\eV \per \angstrom}.
Lastly, a plane-wave energy cutoff of \SI{400}{\eV} was found to sufficiently minimize the total energy when used with a Gaussian smearing method and a smearing width of \SI{0.01}{\eV}.

\section{Experimental Results} \label{exp}
\subsection{Microtructural Analysis} \label{exp:Struct}
\begin{figure}[t]
\includegraphics[width=\columnwidth]{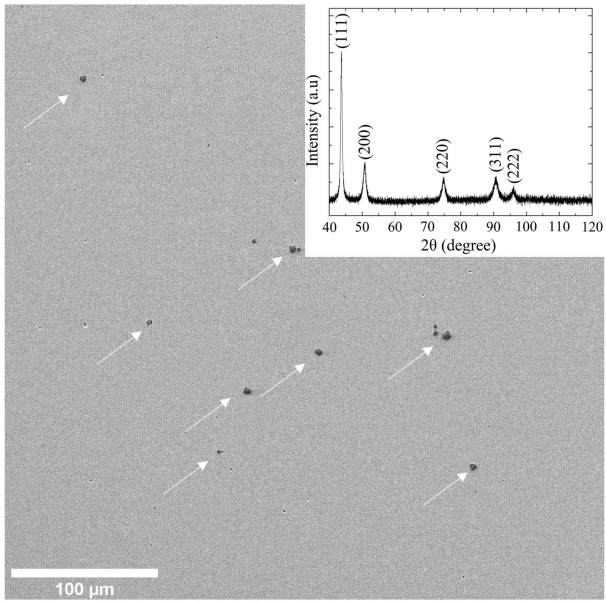}
   \caption{SEM image of homogenized Cantor alloy (anneal A). A small number of particles (black spots identified by arrows) were determined by EDX to be oxide rich in Cr and Mn. The XRD data indicated the sample is single-phase and all peaks could be indexed as FCC.}
   \label{fig:Microstructure SEM and XRD}
\end{figure}

A representative example of the microstructure of the Cantor alloy samples is shown in Fig.~\ref{fig:Microstructure SEM and XRD} for the anneal A sample.
The micro-sized pores observed in the SEM image are considered to be casting porosity.
Additionally, EDX analysis identified a small number of particles as oxides rich in Cr and Mn. 
The sample consists of a single FCC phase according to the XRD pattern, and no peak was identified as corresponding to the oxide particles. 
The existence of such (Cr, Mn)-rich particles in as-cast and homogenized Cantor alloys have been reported in previous studies~\cite{pickering_precipitation_2016,otto_decomposition_2016,otto_microstructural_2014}. 
Furthermore, the XRD data did not contain any peaks associated with inclusions \cite{gali_tensile_2013}, which, due to their small volume fraction, are not considered to significantly influence the properties reported later in this work.
Sample composition was confirmed via an EPMA measurement of the as-cast sample, and the results are presented in Table~\ref{tab:Composition}.

\begin{table}[htbp]
     \caption{Composition of the as-cast equiatomic Cantor alloy, as determined by EPMA measurement.}
      \begin{tabular}{c c}
          \hline
          Element & at. \% \\
          \hline
          \ch{Cr} & 20.31\,$\pm$\,1.05 \\
          \ch{Mn} & 19.44\,$\pm$\,2.01 \\
          \ch{Fe} & 20.48\,$\pm$\,1.32 \\
          \ch{Co} & 20.13\,$\pm$\,0.83 \\
          \ch{Ni} & 19.65\,$\pm$\,0.99 \\
          \hline
      \end{tabular}
      \label{tab:Composition}
\end{table}

\subsection{Magnetization of Homogenized CrMnFeCoNi} \label{exp:Mag}

\begin{figure}
\includegraphics[width=\columnwidth]{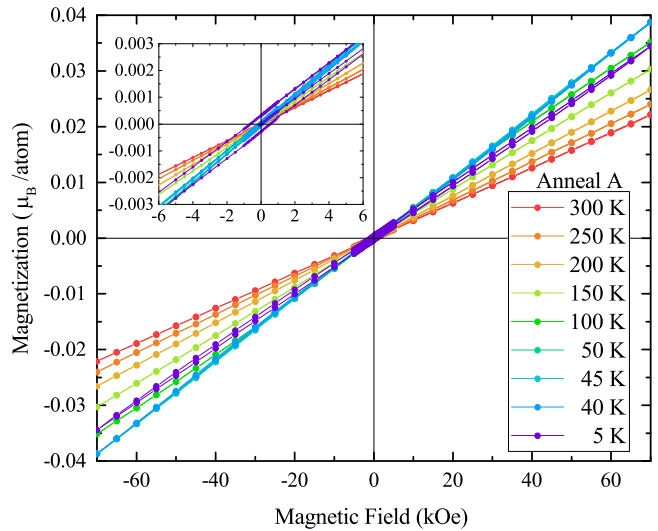}
    \caption{Magnetization versus field data on equiatomic Cantor alloy at a variety of temperatures between \SI{5}{\kelvin} and \SI{300}{\kelvin}.  Inset shows the same data at low field, between \SI{-6}{\kilo \oersted} and \SI{6}{\kilo \oersted}. The magnetization remains small to the highest measured fields.}
    \label{fig:Equi_A_MvsH}
\end{figure}

A magnetization versus temperature plot for a sample of equiatomic Cantor alloy at different temperatures from \SI{5}{\kelvin} to \SI{300}{\kelvin} is displayed in Fig.~\ref{fig:Equi_A_MvsH}.
All of the displayed curves remain roughly linear across the measured range of \SI{-70}{\kilo \oersted} to \SI{70}{\kilo \oersted} with magnitudes of only a few hundredths of a bohr magneton, demonstrating that the compound is far from magnetic saturation.
The only feature that clearly distinguishes one curve from another is the temperature dependence as indicated by different slopes for different temperatures.
The \SI{300}{\kelvin} dataset has the lowest slope, and slope increases as temperature decreases until \SI{40}{\kelvin} and \SI{45}{\kelvin}, which largely overlap one another.
As temperature falls below \SI{40}{\kelvin}, the slope begins to decrease again.

\begin{figure}
\includegraphics[width=\columnwidth]{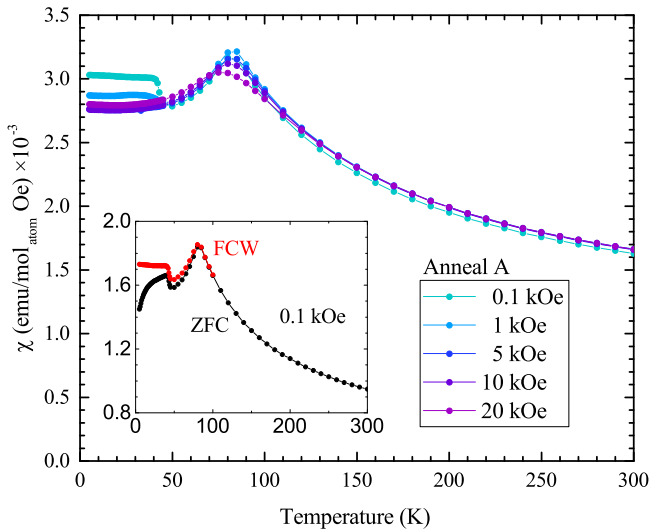}
    \caption{Magnetic susceptibility versus temperature data on equiatomic Cantor alloy under various applied magnetic fields.  The main plot shows only field-cooled warming (FCW) data, while the inset shows compares zero-field cooled (ZFC) and FCW data at \SI{0.1}{\kilo \oersted}.}
    \label{fig:Equi_A_MvsT}
\end{figure}

Field-cooled warming (FCW) magnetization versus temperature data from a sample of anneal A are displayed in Fig.~\ref{fig:Equi_A_MvsT} at fields from \SI{0.1}{\kilo \oersted} to \SI{20}{\kilo \oersted}, while the inset compares both zero-field cooled (ZFC) and FCW data at \SI{0.1}{\kilo \oersted}.
Magnetization plots are presented in units of \si{\emu \per \molatom \per \oersted} to 
simplify estimation of effective moment, $p_{\rm{eff}}$, which can be written as 
\begin{equation}
    p_{\rm{eff}} = 2.82\,C_{m}^{1/2}\;,
    \label{eq:peff}
\end{equation}
where $p_{\rm{eff}}$ is in units of bohr magnetons and $C_m$ is the Curie constant in units of \si{\emu \kelvin \per \molatom \per \oersted}~\cite{mcelfresh_fundamentals_1994}.
In this notation, \si{\molatom} denotes moles of atoms, as opposed to, for example, moles of formula units or moles of magnetic ions, and \si{\emu} is defined as \si{\erg \per \gauss}, as reported by the Quantum Design MPMS.
At \SI{0.1}{\kilo \oersted}, two transitions are clearly seen:  a step-like anomaly at \SI{43}{\kelvin}, and a peak at \SI{85}{\kelvin}.
As field increases, the \SI{43}{\kelvin} transition is rapidly suppressed, becoming notably smaller at \SI{1}{\kilo \oersted} and vanishing completely at \SI{5}{\kilo \oersted} and above.
The higher temperature transition is less affected by the increase in field.
At an applied field of \SI{20}{\kilo \oersted}, the higher temperature transition is slightly smaller, slightly broadened and shifted down in temperature from \SI{85}{\kelvin} to \SI{75}{\kelvin} compared to the \SI{0.1}{\kilo \oersted} curve.
The small size of the \SI{43}{\kelvin} transition is notable, which led us to examine the possibility that it could derive from one the known impurity phases~\cite{schuh_mechanical_2015,otto_decomposition_2016,li_accelerated_2018}.

\subsection{Effects of Plastic Deformation} \label{exp:Squish}
\begin{figure}
\includegraphics[width=\columnwidth]{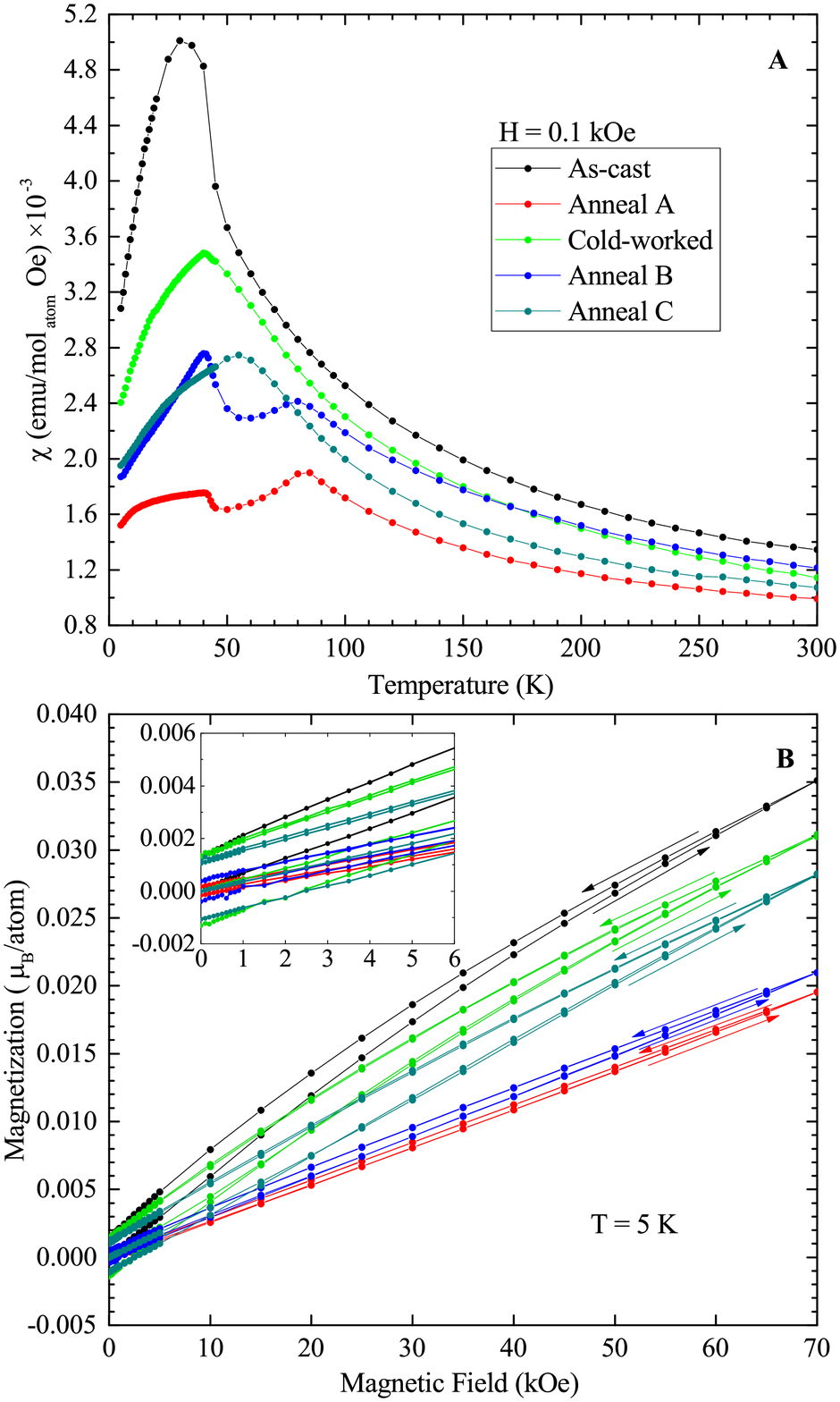}
    \caption{Effects of annealing and cold working on the magnetic properties of equiatomic Cantor alloy.  (A) ZFC susceptibility as a function of temperature at \SI{0.1}{\kilo \oersted}.  (B) Magnetization as a function of temperature at \SI{5}{\kelvin}.  Data are plotted in units of bohr magnetons per atom to show that the compounds are not close to saturation across the range of measured magnetic field. Larger hysteresis in the cold-worked and anneal C samples may be indicative of ferromagnetic precipitates.  Inset shows the same datasets at low field, up to \SI{6}{\kilo \oersted}.}
    \label{fig:ProcessingComparison}
\end{figure}

Since the Cantor alloy is known to form precipitates as a result of deformations and moderate temperature anneals~\cite{schuh_mechanical_2015,otto_decomposition_2016,li_accelerated_2018}, investigation of the processing history dependence of the compound is necessary in order to understand the origin of the magnetic behavior.
The \SI{43}{\kelvin} transition in particular is notably small, with step size of approximately $\mathrm{1 \times 10^{-5}\ \mu_B\ atom^{-1}}$ in fields of \SI{0.1}{\kilo \oersted} and \SI{1}{\kilo \oersted}, suggesting that it may originate from a magnetic secondary phase which comprises a proportionally small amount of the sample.
This hypothesis is tested through magnetization measurements performed on samples which have undergone different cold-working and annealing processes.
The data are plotted in Fig.~\ref{fig:ProcessingComparison}, in which each curve represents a different processing history, as described in the Methods section.
In Fig.~\ref{fig:ProcessingComparison}A, the as-cast sample shows a large \SI{43}{\kelvin} transition relative to the other samples, but notably lacks a peak at higher temperature.
Annealing the as-cast material at \SI{1100}{\celsius} produces the anneal A sample, in which the \SI{45}{\kelvin} signature is significantly smaller and sharper than the corresponding feature in the as-cast sample.
Additionally, a sharp peak appears at \SI{85}{\kelvin} in the anneal A data.
Further changes of the magnetic response are observed after cold-working the anneal A sample in a hydraulic press (cold-worked), and subsequently, annealing at \SI{700}{\celsius} for one hour (anneal B), and at \SI{1100}{\celsius} for six days (anneal C).
These datasets demonstrate that cold-working reduces the size of the \SI{43}{\kelvin} transition and pushes the higher temperature transition to lower temperatures while increasing its size.
However, annealing, even for a brief duration, results in a \SI{43}{\kelvin} transition larger than what appears in the cold-worked samples but smaller than that of the arc-melted as-cast.
In addition, in moving from the as-cast to the annealed samples, the higher temperature transition is revealed, even if it remains smaller than in the cold-worked sample.
This strong processing dependence also explains differences between the results of our measurements and those of others~\cite{jin_tailoring_2016,schneeweiss_magnetic_2017,kamarad_effect_2019}.

Magnetization versus magnetic field data for the same samples are shown in Fig.~\ref{fig:ProcessingComparison}B, using the same color-coding as Fig.~\ref{fig:ProcessingComparison}A.
The data are plotted in units of bohr magnetons per atom to clearly demonstrate that none of the measured samples are close to magnetic saturation up to fields of \SI{70}{\kilo \oersted}.
Furthermore, these datasets show the magnetization of processed samples are similar in magnitude to those of the `pristine' annealed sample.
The inset plot shows a detailed view of low-field data below \SI{6}{\kilo \oersted}.
Anneal A reveals the lowest level of hysteresis, while the cold-worked and anneal C samples show the highest.
Larger hysteresis may be indicative of the emergence of ferromagnetic secondary phases.

\subsection{Impurity Contribution to Magnetic Properties} \label{exp:Imp}
\begin{figure}
\includegraphics[width=\columnwidth]{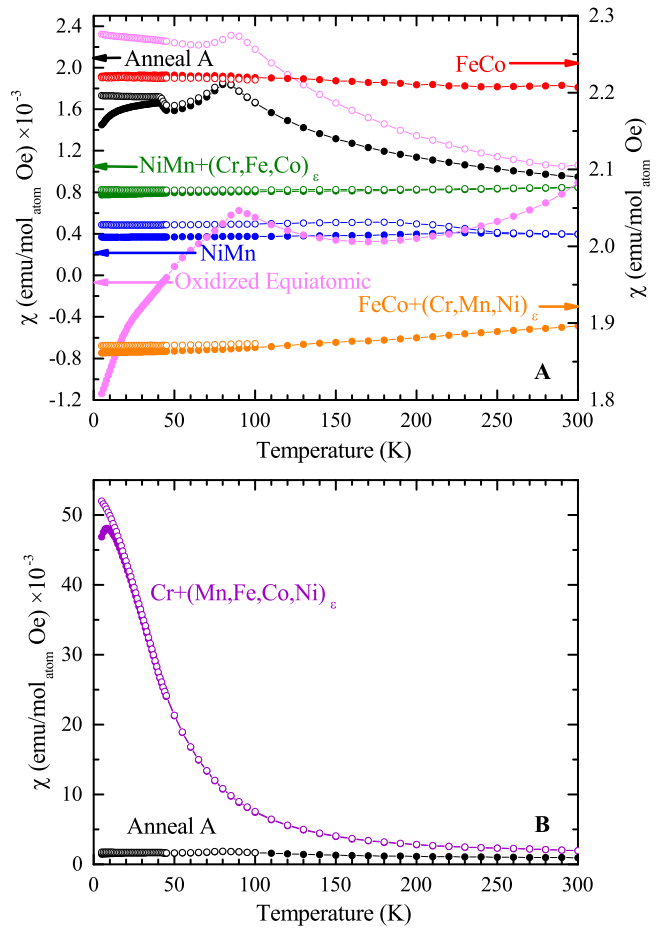}
    \caption{(A) Comparison of magnetic susceptibility versus temperature data from FeCo and NiMn precipitate compounds as well as an oxidized equiatomic compound and anneal A.  
    FeCo-based compounds exhibit comparatively large magnetization and are measured on the right y-axis for clarity.  
    (B)  Comparison of magnetic susceptibility data from the Cr-rich precipitate compound and anneal A.  
    The  ``+(Cr,Fe,Co)$_\epsilon$,'' ``+(Cr,Mn,Ni)$_\epsilon$'' and ``+(Mn,Fe,Co,Ni)$_\epsilon$''  suffixes indicate the presence of small amounts of other Cantor alloy elements consistent with the precipitate compositions observed in the works of 
    Otto \emph{et al.}~\cite{otto_decomposition_2016}, 
    Schuh \emph{et al.}~\cite{schuh_mechanical_2015}, and Li \emph{et al.}~\cite{li_accelerated_2018}.  Closed circles represent zero-field cooled (ZFC) data, while open circles represent field-cooled warming (FCW) data. All data is measured in emu per moles of atoms per Oe.}
    \label{fig:precipitates}
\end{figure}

Having investigated the effects of different processing steps on the equiatomic Cantor alloy itself, the next step was to synthesize and measure the magnetism of the precipitates, shown in Fig.~\ref{fig:precipitates}, with the annealed equiatomic, NiMn, and NiMn+(Cr,Fe,Co)$_\epsilon$ on the left y-axis of Fig.~\ref{fig:precipitates}A and FeCo and FeCo+(Cr,Mn,Ni)$_\epsilon$ on the right axis.
Despite synthesizing and annealing in Ar atmosphere, oxidation of the alloy is also considered, and data from the oxidized sample are also included on the left axis of pannel A.
Measurements performed on the \ch{Cr}-rich precipitate reveal magnetization values significantly higher than those of the \ch{NiMn} compounds and significantly lower than the \ch{FeCo} compounds, such that the \ch{Cr} compound data do not fit easily alongside either.
For clarity, this curve has been separated into pannel B, where it is compared to anneal A independently.
The suffixes of ``+(Cr,Fe,Co)$_\epsilon$,'' ``+(Cr,Mn,Ni)$_\epsilon$'' and ``+(Mn,Fe,Co,Ni)$_\epsilon$'' indicate the presence of small amounts of other Cantor alloy elements added to match the compositions of precipitates reported by Otto \emph{et al.}~\cite{otto_decomposition_2016}, Schuh \emph{et al.}~\cite{schuh_mechanical_2015}, and Li \emph{et al.}~\cite{li_accelerated_2018}.
The investigated precipitates lack any signs of a transition near \SI{43}{\kelvin}, suggesting that this transition is, in fact, an intrinsic aspect of the Cantor alloy system.
While the \SI{85}{\kelvin} transition is present in the oxidized curve, its size is not significantly changed relative to the anneal A curve.
Furthermore, this transition is also not present in any of the precipitate curves, suggesting that it is also intrinsic to the Cantor alloy.

\subsection{Muon Spin Relaxation} \label{exp:Muon}
\begin{figure}
	\includegraphics[width=\columnwidth]{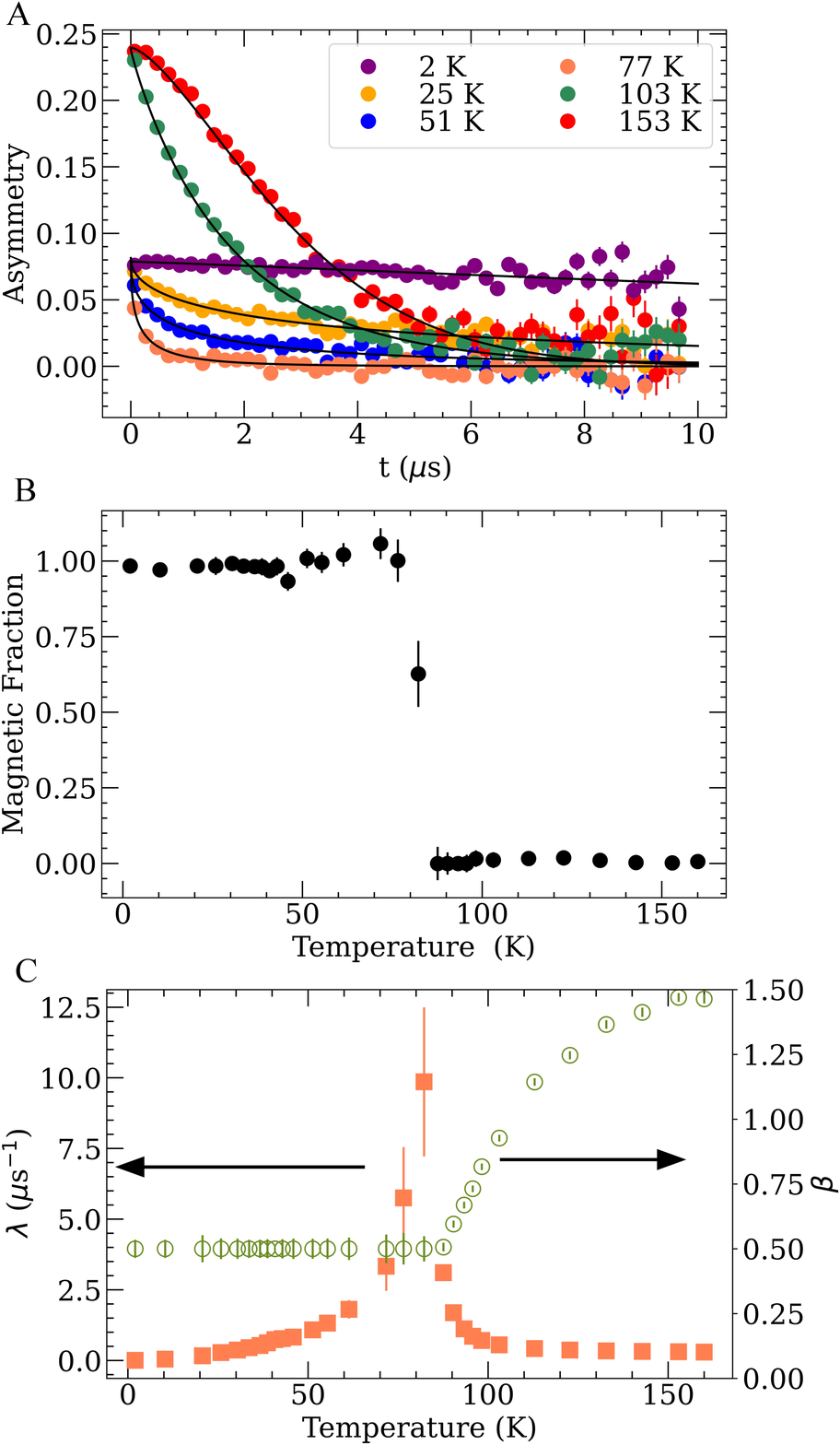}
	\caption{\label{fig:muSR} (a) Representative \muSR\ asymmetry spectra collected at various temperatures in zero field. The colored symbols show the data, and the black solid curves show the fits described in the main text. (b) Temperature dependence of the volume fraction of the sample exhibiting static magnetism. (c) Temperature dependence of the relaxation rate $\lambda$ (left vertical axis) and relaxation exponent $\beta$ (right vertical axis) determined from fits to the asymmetry spectra.
	}		
\end{figure}

Key experimental results from the \muSR\ measurements of a sample from anneal A are shown in Fig.~\ref{fig:muSR}.
Panel A displays asymmetry spectra collected at representative temperatures between 2~K and 153~K.  
At the highest temperature studied, a gentle relaxation with a Gaussian-like nature is observed.  
Applying a field of 1~kOe parallel to the initial muon spin polarization recovers approximately 75\% of the asymmetry (data not shown), indicating that the relaxation at this temperature is due primarily to weak, random dipolar fields from nuclear moments~\cite{uemur;ms99}. 
As the temperature is lowered, the relaxation increases and becomes more exponential in nature, characteristic of magnetic fluctuations from electronic spins. 
Between 87~K and 82~K, the initial asymmetry drops rapidly from $\sim$0.24 to $\sim$0.135 and then further to 0.08 (i.e, 1/3 the total initial asymmetry) at 77~K. 
This ``1/3 tail'' is the hallmark of a polycrystalline material with static magnetism throughout the full sample volume. 
No coherent oscillations of the asymmetry are observed in the ``2/3 component'', indicating very large fields and/or a broad distribution of field strengths at the muon stopping sites~\cite{uemur;ms99}. 
As the temperature is lowered further, the relaxation of the 1/3 component gradually slows until almost no relaxation remains at 2~K, consistent with magnetic fluctuations freezing out. 
Qualitative inspection of the spectra therefore confirms the presence of a sharp and uniform transition throughout the full sample volume with an onset temperature between 82 and 87~K. 
No pronounced change in the spectra is observed around 43~K, which demonstrates the transition at this temperature cannot be due simply to a minority phase in the sample, otherwise a further drop of the asymmetry would be observed below 43~K.

To gain more quantitative insight, least-squares fits to the asymmetry data were performed using the stretched exponential function $a(t) = a_0 e^{(-\lambda t)^{\beta}}$, where $a(T)$ is the time-dependent asymmetry, $a_0$ is the initial asymmetry at $t=0$, $\lambda$ is the relaxation rate, and $\beta$ is the exponential power. 
This type of stretched exponential function is phenomenological, employed to model a continuous distribution of relaxation rates~\cite{crook;jpcm97}.
The best-fit asymmetry curves agree well with the data, as seen by the solid curves in Fig.~\ref{fig:muSR}A. 
The volume fraction of the sample that exhibits static magnetism can be determined from the initial asymmetry values as $f(T) = \left( a_0^{\mathrm{max}} - a_0(T) \right) / \left( a_0^{\mathrm{max}} - a_0^{\mathrm{LT}} \right)$, where $a_0(T)$ is the best-fit value of $a_0$ at temperature $T$, and $a_0^{\mathrm{max}}$ is the maximum value of $a_0$ across all measured temperatures, and $a_0^{\mathrm{LT}}$ is the average value of $a_0$ for the low-temperature data ($T<70$~K). 
As seen in Fig.~\ref{fig:muSR}B, $f(T)$ transitions rapidly from 0 to 1 in a small temperature window centered around 82~K, indicating that the sample undergoes a highly uniform magnetic transition. 
Evidence of this transion is also shown in Fig.~\ref{fig:muSR}C, which displays the relaxation rate $\lambda$ and the exponential power $\beta$ as a function of temperature. 
The prominent peak in $\lambda$ centered around 82~K is evidence of critical spin dynamics as the temperature decreases toward the transition, as is observed in canonical spin glasses and continuous phase transitions~\cite{uemur;prb85,yaoua;b;msr11}. 
The exponential power $\beta$ is $\sim$1.5 at high temperature where relaxation from nuclear dipolar fields dominates, but it decreases steadily as the temperature is lowered and electronic spin fluctuations become more prominent. 
At the transition temperature and below,  $\beta$ converges to values between 0.45 and 0.55 when left fully unconstrained, and for consistency, $\beta$ was set to 0.5 for the spectra collected at 82~K and below. 
This value of $\beta$ is expected when the system exhibits multiple relaxation channels and/or spin fluctuation rates~\cite{dunsi;prb96}, which is unsurprising in this highly disordered alloy.
Similar values were also observed in the entropy-stabilized antiferromagnetic oxide (Mg,Co,Ni,Cu,Zn)O~\cite{frand;prm20}.

\subsection{Specific Heat} \label{exp:HC}
\begin{figure}
\includegraphics[width=\columnwidth]{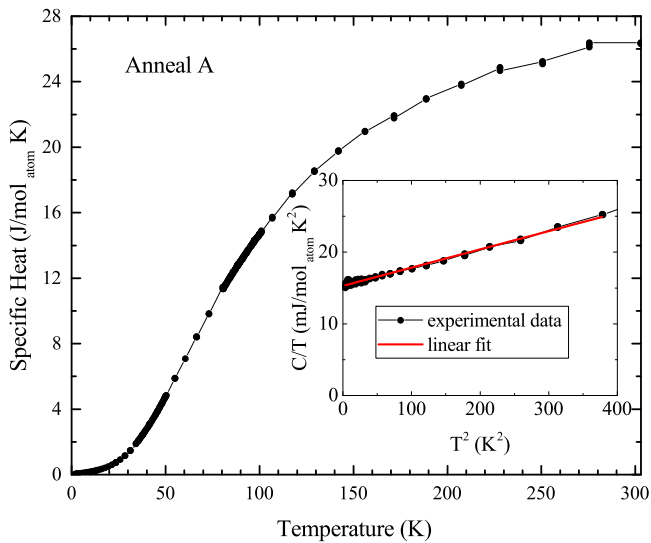}
    \caption{Specific heat as a function of temperature for equiatomic Cantor alloy sample anneal A.  Inset shows a plot of specific heat divided by temperature versus temperature squared.}
    \label{fig:HC}
\end{figure}

Specific heat data are used in later sections to calculate magnetic entropy and the effective mass of charge carriers in order to better understand the nature of the magnetism present in the Cantor alloy.
Results of specific heat versus temperature measurements on the anneal A sample are shown in Fig.~\ref{fig:HC}.
Data collection focused on the regions around \SI{43}{\kelvin} and \SI{85}{\kelvin}, but no features are apparent in this plot despite the transitions visible in the magnetization data at these temperatures.
The inset contains a plot of $C/T$ versus $T^2$, focusing on the low-temperature region.
The linear fit included in the inset follows from 
\begin{equation}
    C = \gamma T + \beta_{\mathrm{ph}} T^3\;,
    \label{eq:HC}
\end{equation}
the expression for the low-temperature specific heat of a metal~\cite{tari_specific_2003}.
In this equation, the $\gamma T$ term represents the electronic component of the specific heat, while $\beta_{\mathrm{ph}} T^3$ is the lattice specific heat and is related to the Debye temperature.
The results of fitting this equation are
\begin{align*}
\gamma\mathrm{\ = 15.3(2)\ mJ\ mol^{-1}_{atom}\,K^{-2}} \\
\beta_{\mathrm{ph}}\mathrm{\ = 0.025(3)\ mJ\ mol^{-1}_{atom}\,K^{-4}}.
\end{align*}

\subsection{Hall Effect and Carrier Density} \label{exp:Hall}
\begin{figure}
\includegraphics[width=\columnwidth]{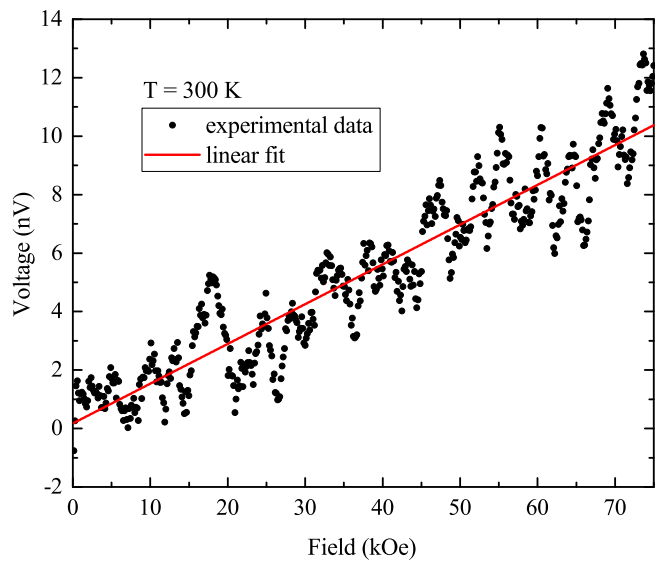}
    \caption{Hall voltage of the equiatomic Cantor alloy sample anneal A.  Data has been anti-symmetrized to remove any $r_{xx}$ contribution.}
    \label{fig:Hall}
\end{figure}
Measurements of Hall voltage as a function of magnetic field permit the estimation of carrier density according to 
\begin{equation}
    V_H = \frac{IB}{ned}\,,
    \label{eq:Hall}
\end{equation}
where \emph{I} is the current, \emph{B} is the magnetic field perpendicular to the current, \emph{e} is the electron charge, and \emph{d} is the sample thickness.
The carrier density can, in turn, provide insight into the magnetism of the alloy as carried out in Section~\ref{disc:meff}.
Results of Hall measurements are shown in Fig.~\ref{fig:Hall}, including a linear fit with a slope of \SI{1.4(4)E-4}{\nano \volt \per \oersted}.
Using this slope in combination with Eq.~\ref{eq:Hall}, a carrier density $n =\ $\SI{2.9(8)E22}{\per \cubic \centi \meter} can be calculated.
This result is consistent with the carrier density of Al$_x$CrFeCoNi reported by Kao \emph{et al.}~\cite{kao_electrical_2011}, as well as with that of a typical metal ($\sim$~\SI{1E22}{\per \cubic \centi \meter} to \SI{1E23}{\per \cubic \centi \meter})~\cite{ashcroft_solid_1976,singleton_band_2001}.

\section{Discussion} \label{disc}
Experimental results reported in this work demonstrate the effects of processing on the two observed transitions in the equiatomic Cantor alloy as well as the intrinsic nature of these transitions.
However, the magnetic properties can be examined in more detail, and information on the nature of magnetism in this sample can be extracted from the presented experimental data.
This section begins with Curie-Weiss analysis of the magnetic susceptibility which reveals a significant constant offset.
Analysis which combines specific heat data, computational results, and Hall measurements points to Stoner enhancement as the origin of the offset.
The Stoner enhancement parameter \emph{Z} is calculated through three different methods which obtain a \emph{Z} value between 0.92 and 0.97, suggesting that this compound is on the cusp of magnetic order.
This result is perhaps unsurprising given the observed magnetic transitions and the strong magnetic ordering seen in the Cantor alloy's constituent elements, but more notable is the small size of the magnetic transitions, which may be due to the presence of antiferromagnetic interactions, as evidenced by DFT results.
These findings are consistent with previous work on a similar compound, \ch{Fe40Mn40Cr10Co10}~\cite{egilmez_magnetic_2021}.

\subsection{Modified Curie-Weiss Fitting} \label{disc:CW}
\begin{figure}
\includegraphics[width=\columnwidth]{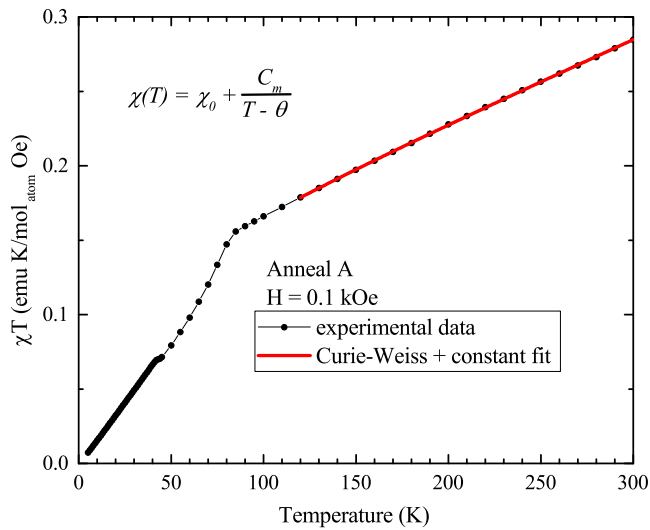}
    \caption{Susceptibility times temperature versus temperature plot for equiatomic Cantor alloy, alongside a modified Curie-Weiss plus constant fit.  
    The two transitions can be observed as sharp changes in slope.  
    Above the \SI{85}{\kelvin} transition, the plot is roughly linear with a positive slope.}
    \label{fig:Equi_A_ChiTFit}
\end{figure}

When using a Curie-Weiss expression to estimate the effective moment of the anneal A sample at temperatures above the \SI{85}{\kelvin} transition, reasonable fits required a modified expression written as
\begin{equation}
    \chi(T) = \chi_{\rm{o}} + \frac{C_{m}}{T-\theta}\;,
    \label{eq:modCW}
\end{equation}
where $C_m$ is the Curie constant, $\theta$ is the Curie temperature, and $\chi_{\rm{o}}$ is a temperature-independent constant.
This dataset is plotted as susceptibility times temperature for greater readability across the entire temperature range.
The resulting fit is compared with experimental data in Fig.~\ref{fig:Equi_A_ChiTFit}, and the resultant value for each parameter is listed in Table~\ref{tab:suscFitValues}.

\begin{table}[htbp]
    \caption{Constant offset $\chi_{\rm{o}}$, Curie constant $C_m$, Curie temperature $\theta$, and effective moment $p_{\rm{eff}}$ from modified Curie-Weiss fit, Eq.~\ref{eq:modCW}.  The value of $p_{\rm{eff}}$ was calculated using Eq.~\ref{eq:peff} and the value of $C_m$.}
    \begin{tabular}{c c c}
        \hline
        Variable & Value & Units\\
        \hline
         $\chi_{\rm{o}}$ & $\mathrm{5.4 \times 10^{-4}}$ & $\mathrm{emu\ mol^{-1}_{atom} Oe^{-1}}$ \\
         $C_m$ & 0.13 & $\mathrm{emu\ K\ mol^{-1}_{atom} Oe^{-1}}$ \\
         $\theta$ & -16.1 & K \\
         $p_{\rm{eff}}$ & 1.01 & $\mathrm{\mu_B\ atom^{-1}}$ \\
         \hline
    \end{tabular}
    \label{tab:suscFitValues}
\end{table}

This fitting equation and its results differ notably from those of Kamar\'ad \emph{et al.},~\cite{kamarad_effect_2019} who used a non-modified Curie-Weiss equation to obtain a Curie temperature of $\theta=\,$\SI{-210}{\kelvin} and an effective moment of $p_{\rm{eff}} = 2.71\,\mathrm{\mu_B}$/f.u., using their formula of Cr$_{0.205}$Mn$_{0.20}$Fe$_{0.205}$Co$_{0.199}$Ni$_{0.191}$.
Given the strong history dependence of the Cantor alloy, this discrepancy may be a result of the differing processing methods, since Kamar\'ad \emph{et al.} cold-rolled their samples after casting and then annealed at \SI{1173}{\kelvin} for \SI{1}{\hour}.
Our investigation of Cantor alloy processing history dependence shows that cold-working will increase magnetization, and subsequent high-temperature annealing will not restore the sample to a ``pristine'' state.

\subsection{Specific Heat and Entropy} \label{disc:HC}
The fit in Fig.~\ref{fig:HC} provides the parameter $\beta$, which is the coefficient of the phonon contribution to the specific heat.
In the low-temperature limit, $\beta$ can be used to calculate the Debye temperature $\Theta_D$ according to 
\begin{equation}
    \beta = \frac{12\pi^4}{5}\frac{N_Ak_B}{\Theta_D^3}\,,
    \label{eq:HC_beta}
\end{equation}
in which $N_A$ is Avogadro's number~\cite{schroeder_introduction_2000}.
This calculation gives $\Theta_D$ = \SI{427}{\kelvin}, not far from the Debye temperatures of some of the Cantor alloy's component elements~\cite{kittel_introduction_2005}.
The Debye temperature can also be estimated by visual comparison of the phononic component of the specific heat to the Debye function, plotted as $C$ versus $T/\Theta_D$~\cite{tari_specific_2003}.
If $\Theta_D$ is chosen correctly, the two curves should closely align.
The phononic specific heat is estimated by subtracting $\gamma T$ from the experimentally obtained specific heat, with the value of $\gamma$ taken from the fit shown in the inset of Fig.~\ref{fig:HC}.
This process produces a Debye temperature $\Theta_D$ = \SI{380}{\kelvin}.
These values are displayed in Table~\ref{tab:Debye} for comparison.

\begin{table}[htbp]
    \caption{Debye temperatures of the equiatomic Cantor alloy and its component elements.  Elemental Debye temperatures are obtained from Kittel~\cite{kittel_introduction_2005}.}
    \begin{tabular}{c c}
        \hline
        Compound & $\Theta_D$ (K) \\
        \hline
         \ch{CrMnFeCoNi}, Eq.~\ref{eq:HC_beta} & 427 \\
         \ch{CrMnFeCoNi}, Debye function comparison & 380 \\
         \ch{Cr} & 630 \\
         \ch{Mn} & 410 \\
         \ch{Fe} & 470 \\
         \ch{Co} & 445 \\
         \ch{Ni} & 450 \\
         \hline
    \end{tabular}
    \label{tab:Debye}
\end{table}

\begin{figure}
\includegraphics[width=\columnwidth]{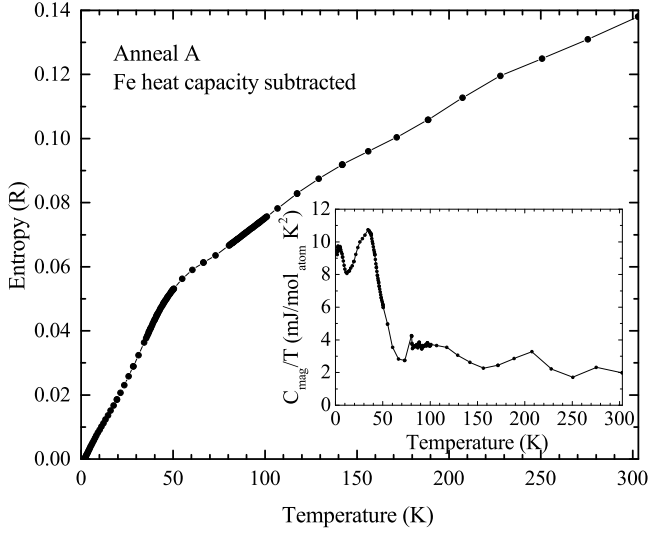}
    \caption{Magnetic entropy as a function of temperature in units of R, the gas constant.  The inset plot shows an estimate of the magnetic component of the specific heat produced by subtracting rescaled \ch{Fe} specific heat data from the Cantor alloy specific heat data shown in Fig.~\ref{fig:HC}.}
    \label{fig:HC_entropy}
\end{figure}

An estimate of the magnetic entropy of the equiatomic Cantor alloy is shown in Fig.~\ref{fig:HC_entropy}, while the inset contains the magnetic component of the specific heat divided by temperature as a function of temperature.
In order to determine the magnetic contribution, the specific heat of Fe as a function of temperature was compiled from multiple sources~\cite{austin_heat_1932,keesom_atomic_1939,kelley_specific_1943} in the assumption that it would approximate the phononic component of the specific heat in the Cantor alloy.
The \ch{Fe} data were smoothed, then rescaled by a factor of 403/470 to account for differences in Debye temperatures between the equiatomic Cantor alloy and elemental \ch{Fe}.
Since two different values of the Cantor alloy Debye temperature are obtained through different methods, this ratio uses the average of these values.
The re-scaled fit data were subtracted from the specific heat data shown in Fig.~\ref{fig:HC}, and the results are shown in Fig.~\ref{fig:HC_entropy} in the form of $C_{\rm{mag}}/T$ versus $T$.
Integrating over the magnetic specific heat divided by temperature gives the entropy associated with magnetism in the Cantor alloy.

\subsection{Effective Mass and Carrier Density} \label{disc:meff}
\begin{figure}
\includegraphics[width=3.3in]{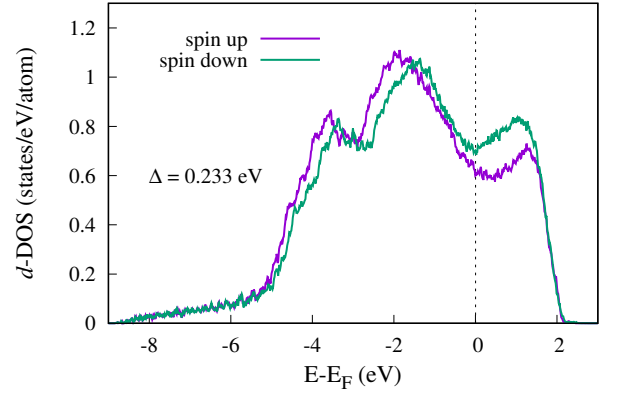}
    \caption{The \textit{d}-projected total density of states calculated from DFT. 
    The spin-up states are shown by the purple line while spin-down states are shown using green. 
    The $\Delta$ value is the exchange splitting parameter needed to evaluate the Stoner Criterion and represents the splitting in energy between the up and down states. 
    This result is calculated by measuring the average separation (in eV) of the identifiable peaks and features in the occupied region of the DOS (\textit{i.e.}, below the Fermi level). 
    These DOS curves represent a combination of the individual DOS plots from each DFT simulation, normalized for the total number of atoms.}
    \label{fig:dft_dos}
\end{figure}

The fit shown in Fig.~\ref{fig:HC} to the specific heat also provides an estimation of the electronic contribution through the coefficient $\gamma$.
This value can be used to calculate the effective carrier mass $m^*$ according to
\begin{equation}
    \frac{m^*}{m_e} = \frac{\gamma}{\gamma_0},
    \label{eq:effmass}
\end{equation}
in which $m_e$ is the mass of a free electron~\cite{tari_specific_2003}.
The variable $\gamma_0$ is a theoretical value calculated from the electronic DOS and can be expressed as
\begin{equation}
    \gamma_0 = (\pi k_B)^2\frac{g(E_F)}{3}\;,
    \label{eq:gammaDOS}
\end{equation}
in which $k_B$ is the Boltzmann constant and $g(E_F)$ is the DOS in the \textit{d}-band at the Fermi level, $E_F$~\cite{ziman_principles_1972}. In metals such as the Cantor alloy studied here, the \textit{s}- and \textit{p}-bands are broad near the Fermi level and thus contribute very little to the overall electron occupation. In other words, only the \textit{d}-band needs be taken into account~\cite{Hammer1995}.
The spin separated projected DOS of the \textit{d}-band, taken from the combined DOS of the four DFT calculations in this work, and normalized to a per atom occupation is presented in Fig.~\ref{fig:dft_dos}.
The spin-up and spin-down occupations at the Fermi energy are $\mathrm{0.613 \ states \ eV^{-1} atom^{-1}}$ and $\mathrm{0.736 \ states \ eV^{-1} atom^{-1}}$, respectively, resulting in a combined total Fermi level occupation of $g(E_F)=\mathrm{1.349 \ states \ eV^{-1} atom^{-1}}$.
Using this total Fermi level into Eq.~\ref{eq:gammaDOS}, one finds $\gamma_0\mathrm{\ =\ 3.18(9)\ mJ\ mol_{atom}^{-1}\ K^{-2}}$, leading to an effective carrier mass of $m^*\mathrm{\ =\ 4.8(1)}\,m_e$.

This analysis can be extended by first examining the results provided by the Hall studies which provide a carrier density $n = $~\SI{2.9(8)E22}{\per \cubic \centi \meter}.
Combined with a value of the effective mass $m^*$ extracted from the specific heat, estimates of the Fermi energy 
\begin{equation}
    E_F = \frac{\hbar^2}{2m^*}(3\pi^2n)^{2/3}\;,
\end{equation}
and the Pauli paramagnetic susceptibility
\begin{equation}
    \chi_{\rm{P}} = \frac{3n\mu_0\mu_B^2}{2E_F}\;,
    \label{eq:Pauli_n}
\end{equation}
yield $E_F = 0.72(8)\ \mathrm{eV}$ and $\mathrm{2.3(2) \times 10^{-5}\, emu\ mol^{-1}_{atom} Oe^{-1}}$, respectively.
By comparison, fitting experimental susceptibility data yielded a constant offset of $\mathrm{5.4 \times 10^{-4}\, emu\ mol^{-1}_{atom} Oe^{-1}}$, as seen in Table~\ref{tab:suscFitValues}.
A diamagnetic contribution to the temperature independent part of the susceptibility is also expected, consisting of Landau diamagnetic and core diamagnetic components~\cite{mulay_theory_1976}.
Landau diamagnetic susceptibility can be determined by
\begin{equation}
    \chi_L = -\frac{1}{3}\frac{m_e^2}{m^{*2}}\chi_P\;,
\end{equation}
which produces a value of $\chi_L = \mathrm{-3.3(3) \times 10^{-7}\, emu\ mol^{-1}_{atom} Oe^{-1}}$~\cite{mulay_theory_1976}.
The core diamagnetism depends on the ionization of the atoms within the compound~\cite{bain_diamagnetic_2008}, which is currently not known.
However, based on the possible ionization states of \ch{Cr}, \ch{Mn}, \ch{Fe}, \ch{Co}, and \ch{Ni}, the core diamagnetic susceptibility $\chi_{\rm{core}}$ must be between $\mathrm{-6.0 \times 10^{-7}\, emu\ mol^{-1}_{atom} Oe^{-1}}$ and $\mathrm{-1.1 \times 10^{-6}\, emu\ mol^{-1}_{atom} Oe^{-1}}$.
These estimates suggest that the effects of Landau diamagnetism are negligible in the overall susceptibility of the equiatomic Cantor alloy, and the core diamagnetism may account for at most a 5\% reduction in the temperature-independent susceptibility.

A value for the Pauli paramagnetic susceptibility can also be determined from first principles methods according to 
\cite{blundell_magnetism_2001}
\begin{equation}
    \chi_{\rm{P}} = \mu_0\mu_B^2g(E_F)\;.
    \label{eq:PauliSusc}
\end{equation}
This equation gives a Pauli paramagnetic susceptiblity of $\mathrm{4.4 \times 10^{-5} \, emu\ mol^{-1}_{atom} Oe^{-1}}$, similar to the value calculated from Hall data.

A number of factors point toward the possibility of Stoner enhanced paramagnetism as the source of the discrepancy between the calculated Pauli susceptibility and the experimentally determined offset.
Firstly, more than half of the component elements of the \ch{CrMnFeCoNi} alloy are known to be ferromagnetic, and the work of Schneeweiss \emph{et al.}~\cite{schneeweiss_magnetic_2017} on this compound identify the \SI{43}{\kelvin} transition as ferromagnetic, though this claim is disputed by Kamar\'ad \emph{et al.}~\cite{kamarad_effect_2019}.
These factors suggest a proximity to magnetic order.
Computational results also show a discrepancy in the spin up and spin down DOS, which indicates a spontaneous splitting of energy states.
Alongside a molecular field of sufficient strength, this splitting will result in a Stoner enhancement~\cite{blundell_magnetism_2001}.

\subsection{Stoner Enhancement of Pauli Paramagnetism} \label{disc:StonP}
The static component of the susceptibility is given by
\begin{equation}
    \chi_{\rm{static}} = \frac{\chi_P}{1 - Z}\;,
    \label{eq:StonerEnh}
\end{equation}
in which \emph{Z} is the Stoner enhancement parameter, a value between zero and one~\cite{blundell_magnetism_2001,jia_nearly_2007}.
Assuming that $\chi_{\rm{o}}$ from the modified Curie-Weiss fit is an accurate measure of the static component of the susceptibility, the Pauli paramagnetism can be used to calculate \emph{Z}.
The $\chi_P$ obtained from Hall data gives a value of \emph{Z} = 0.96.
Calculating the Pauli susceptibility from the computational DOS using Eq.~\ref{eq:PauliSusc} produces a Stoner enhancement of \emph{Z} = 0.92.
Alternatively, \emph{Z} can be calculated using $\gamma$, the linear component of the specific heat using Eq.~\ref{eq:gammaDOS}, in combination with
\begin{equation}
    \chi_P = \mu_0\mu_B^2g(E_F)\;~\cite{blundell_magnetism_2001},
    \label{eq:chiP_DOS}
\end{equation}
to obtain an equation for the Stoner enhancement parameter \emph{Z} as a function of $\gamma$, \emph{i.e.}
\begin{equation}
    Z = 1 - \frac{3\mu_B^2}{\pi^2k_B^2}\frac{\gamma}{\chi_{\rm{static}}}\;~\cite{jia_nearly_2007}.
    \label{eq:finalZ}
\end{equation}
Using the value of $\gamma$ obtained from specific heat measurements, this expression gives a value of \emph{Z} = 0.97.
These Stoner enhancement parameters are listed alongside those of other known compounds listed in Table~\ref{tab:StonerParam}.
\begin{table}[tbp]
    \caption{Comparison of experimentally determined Stoner enhancement parameters for multiple compounds.  The Stoner parameter calculated from Eq.~\ref{eq:Pauli_n} remains at approximately 0.96 regardless of whether the core diamagnetic correction is applied.}\label{tab:StonerParam}
    \begin{tabular}{c c c}
        \hline
        Material & \emph{Z} & Ref. \\
        \hline
        \ch{CrMnFeCoNi}, Eq.~\ref{eq:Pauli_n} & 0.96 & This work \\
        \ch{CrMnFeCoNi}, Eq.~\ref{eq:PauliSusc} & 0.92 & This work \\
        \ch{CrMnFeCoNi}, Eq.~\ref{eq:finalZ} & 0.97 & This work \\
        \ch{Pd} & 0.82 & \cite{hoare_thermal_1953,chouteau_specific_1968,hong_enhancement_1995,shen_stoner-enhanced_2016}\\
        \ch{TiBe_2} & 0.91 & \cite{stewart_specific_1982}\\
        \ch{HfZn_2} & 0.79 & \cite{radousky_magnetic_1983} \\
        \ch{YFe_2Zn_20} & 0.94 & \cite{jia_nearly_2007} \\
        \ch{YCo_2Zn_20} & 0.50 & \cite{jia_nearly_2007}\\
        \ch{WB_4} & 0.93 & \cite{shen_stoner-enhanced_2016} \\
        \hline
    \end{tabular}
\end{table}
Additionally, Leong \emph{et al.}~\cite{leong_effect_2017} report theoretical values for effective Stoner enhancement parameters for a number of \ch{CoCrFeNi}-based high entropy alloys, all of which were found to be either ferromagnetic or close to ferromagnetic ordering.

One may observe a similarity between Eq.~\ref{eq:finalZ} and the Wilson ratio
\begin{equation}
    R_W = \frac{4\pi^2k_B^2}{3\mu_{0}(g_e\mu_{B})^2} \frac{\chi_0}{\gamma}\,,
    \label{eq:WilsonRatio}
\end{equation}
which provides information on the electron-electron correlations in the compound, where a free electron gas has a Wilson ratio of 1 while a higher ratio indicates stronger interactions~\cite{solyom_fundamentals_2009,paramanik_valence_2013}.
The experimental results reported here give a Wilson ratio of 2.6 for the Cantor alloy anneal A sample.
While this $R_W$ is higher than the $R_W$ of 1.2 to 2.2 observed in the strongly correlated electron compound \ch{Sr2RuO4}, much higher Wilson ratios have been recorded~\cite{julian_normal_1999}.
For example, Balents~\cite{balents_spin_2010} calculates a value of $R_W$ as high as 230 in \ch{FeSc2S4} based on previously published data~\cite{buttgen_spin_2006,fritsch_spin_2004}, perhaps due to competition between spin-orbit coupling and magnetic exchange, and proximity of a quantum critical point~\cite{chen_spin-orbital_2009,balents_spin_2010}.
Similarly, Julian \emph{et al.}~\cite{julian_normal_1999} calculate a Wilson ratio of approximately 40 for the nearly ferromagnetic compound \ch{Ni3Ga}, also using published data~\cite{dood_low_1973,schinkel_applicability_1973,hayden_electronic_1986}.

\subsection{Magnetic State According to the Stoner Criterion} \label{disc:DFT}
The Stoner model of itinerant magnetism states the ferromagnetic phase should be favored if the Stoner criterion, $g_0(E_F)\cdot I \geq 1$, is satisfied, where $g_0(E_F)$ is the DOS per atom at the Fermi level in the non-magnetic case (\textit{i.e.,} from a non-spin-polarized calculation). 
The Stoner exchange parameter $I$ describes the splitting between the spin-up and spin-down states in the magnetic phase. 
In ferromagnetic materials, the wavefunctions of the magnetic and non-magnetic state are identical, however, in the magnetic state the eigenvalues $\epsilon$ are shifted by a constant amount $\mp\,\frac{1}{2}IM$, where $M$ is the magnetization~\cite{blundell_magnetism_2001,zeller2006}. Thus the eigenvalues of the up and down spin states can be written as 
\begin{equation}
    \epsilon^{\uparrow}_{\underline{k}v} = \epsilon^{0}_{\underline{k}v} - \frac{1}{2}IM,
\label{eq:spin_dispersion1}
\end{equation}
\begin{equation}
    \epsilon^{\downarrow}_{\underline{k}v} = \epsilon^{0}_{\underline{k}v} + \frac{1}{2}IM,
\label{eq:spin_dispersion2}
\end{equation}
where $k$ and $v$ denote the wavevectors and band indices. 
By subtracting Eq\@.~\ref{eq:spin_dispersion2} from Eq\@.~\ref{eq:spin_dispersion1}, one finds 
$\Delta = (\epsilon^{\downarrow}_{\underline{k}v} -  \epsilon^{\uparrow}_{\underline{k}v}) = IM$, which can be estimated according the splitting of the spin up and down DOS of the magnetic system, measured in eV. 
Although several methods exist for determining the shift between spin-states, there is no exact or designated method, so any Stoner parameter obtained via the DOS should be considered only an approximation.
In this work, the overall splitting was measured according to the average shifts in the approximate peak locations within the $d$-band DOS, yielding $\Delta \approx 0.233~\rm{eV}$.
Combined with a net magnetization of $0.294~\mathrm{\mu_B}$, this 
$\Delta$ value provides a Stoner exchange parameter $I = 0.793$. 
To complete the Stoner criterion, the Fermi level occupation of the alloy in a non-magnetic state is needed. 
To obtain this value, an additional DFT simulation was performed for the equiatomic alloy in which spin-polarization effects were excluded, yielding a $g_0(E_F)= 1.663 ~\rm{states/(eV \cdot atom)}$ for the non-magnetic alloy. 
Evaluation of the Stoner criterion then yields $g_0(E_F)\cdot I = 1.318$, suggesting that a weakly ferromagnetic phase should should be preferred in this system.

\subsection{Magnetic Order} \label{disc:Order}
The ``step'' in the susceptibility data visible around \SI{43}{\kelvin} is roughly \SI{1E-5}{\bohr \per \atom} at \SI{0.1}{\kilo \oersted} and \SI{1}{\kilo \oersted}.
This transition is notable in its small size, even relative to transitions in ``small-moment'' ferromagnets~\cite{murani_critical_1974,yelland_ferromagnetic_2005,saunders_exceedingly_2020}.
A similar transition was observed in the compound \ch{Fe_{40}Mn_{40}Co_{10}Cr_{10}} by Egilmez and Abuzaid~\cite{egilmez_magnetic_2021}, who identified it as ferrimagnetic and suggested that its small size was due to strong antiferromagnetic coupling between atoms.

\begin{figure*}
\centering
\includegraphics[width=6.5in]{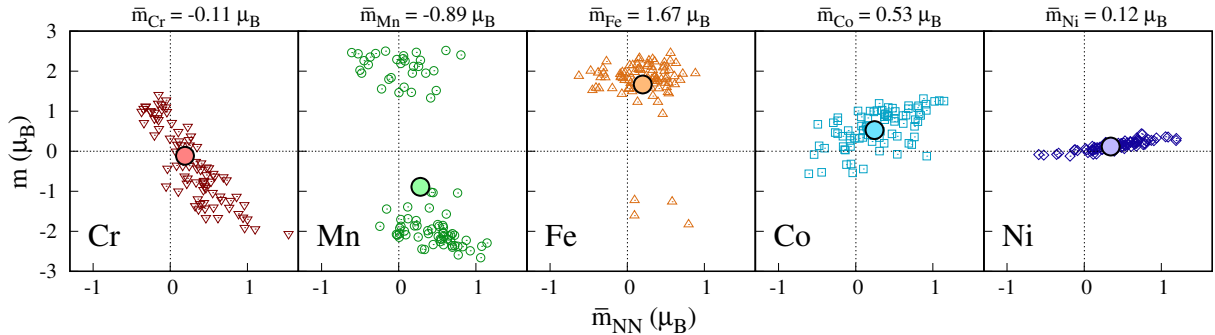}
    \caption{The local magnetic moment of each atom is plotted against the average magnetic moment of its first nearest neighbor shell. The Y-axis depicts the local magnetic moment of each atom (\textit{i.e.,} its spin alignment in the alloy), taken from each DFT simulation. The X-axis shows the average spin alignment of all first nearest neighbor atoms surrounding each atom. Taken together, these show how each atom type prefers to magnetically align with its environment. In the case of ferromagnetic coupling, the atom's spin aligns with its neighbors spins (\textit{i.e.,} Co and Ni). In antiferromagnetic coupling, the atom's spin aligns in the opposite direction of its neighbor's spins (\textit{i.e.,} Cr and Mn). The large black-outlined circle symbols designate the effective ``center of mass" or average of all points for a species. The Y-value of the circle represents the average overall moment of each species type. The X-value of this circle, which is approximately the same for all species, effectively represents the overall magnetic moment of the alloy.
    }
    \label{fig:dft_moments}
\end{figure*}

Insight about the complex and diverse nature of the local magnetic environments is provided by DFT simulations, performed both in this work and in earlier investigations by others studying very similar alloys~\cite{niu2016, niu_magnetically-driven_2018}.
Specifically, a sense of the variations of local magnetism can be garnered from  Fig.~\ref{fig:dft_moments}, where the local atomic magnetic moments, $m$, of each atom in our DFT models are plotted against the average local magnetic moment of their first nearest neighbor shells, $\overline{m}_{\rm{NN}}$.
The atoms are grouped according species type, giving a picture of the type of coupling (\textit{i.e.,} ferro- or antiferromagnetic) each species prefers to form with its local environment.
The total magnetic moment, $p_{\rm{eff}}^{\rm{DFT}}$, from all four simulations, taken as the average moment of all local atomic moments was found to be $0.294~\mathrm{\mu_B}$, which is much lower then the average magnitude (\textit{i.e.,} disregarding spin direction) of the local moments found to be $1.103~\mathrm{\mu_B}$. 
This finding indicates an uneven level of ferro- and anti-ferromagnetic couplings within the alloy, thereby providing a basis for overall (macrocsopic) ferrimagnetic-like ordering.

Examining the couplings exhibited by each species in Fig.~\ref{fig:dft_moments}, one can see that the elements on the left side (\textit{i.e.,} Cr and Mn) of the periodic table \textit{d}-block tend to anti-ferromagnetically align with their local nearest neighbor shell. 
In other words, if the spins of an atom's closest neighbor atoms align on average in one direction, the Cr and Mn atom spins are likely to be found aligned in the opposite direction. 
Conversely, Co and Ni, residing on the right side of the \textit{d}-block, are more likely to align their spins ferromagnetically, or in the same direction, as the spins of their neighbors.
Fe, which sits at the center of the \textit{d}-block, shows neither ferro- nor anti-ferromagnetic tendencies, instead mostly aligning heavily in the spin-up direction regardless of the spins of its neighbors.
It may be that the non-coupled Fe atoms are what ultimately drive the tilt in spin balance towards one direction, away from an overall anti-ferromagnetic order and into the ferrimagnetic regime. 

Experimentally, the Stoner enhancement and the small magnetic moment suggest  that the equiatomic Cantor alloy may be a weak itinerant ferromagnet~\cite{ogawa_magnetic_1967,knapp_ferromagnetism_1971,mattocks_magnetic_1978,vannettE_Field-dependent_2008}.
Furthermore, the magnetic component of entropy shown in the inset of Fig.~\ref{fig:HC_entropy} is approximately 0.13\,R\,ln2 at \SI{43}{\kelvin}, which is an outcome reminiscent of the magnetic entropies of well-known itinerant ferromagnets \ch{ZrZn2} and \ch{Sc3In} (0.02\,R\,ln2 and 0.005\,R\,ln2, respectively~\cite{viswanathan_magnetic_1974,clinton_magnetization_1975,knapp_resistivity_1972}).
This small entropy points toward the involvement of a relatively small number of spins indicative of itinerant magnetism.
Taken together with evidence of antiferromagnetic coupling from DFT calculations and modified Curie-Weiss fitting to experimental data, these pieces of evidence suggest that the equiatomic Cantor alloy is an itinerant ferrimagnet below \SI{43}{\kelvin}.

\section{Conclusion} \label{conc}
By combining a suite of experimental techniques with numerical simulations, this work extends our understanding of the magnetic properties of the equiatomic Cantor alloy CrMnFeCoNi. 
Both experimental and DFT results suggest the presence of weak ferrimagnetism below about \SI{43}{\kelvin}, while \muSR\ measurements indicate a spin-glass-like transition near \SI{85}{\kelvin}. 
In addition, a large offset in the magnetic susceptibility reveals strong Stoner enhancement of the paramagnetism.
Taken with the enhancement of the effective mass and the relatively large Wilson ratio, these results reveal the presence of significant electron-electron interactions within the material.

The transitions observed at \SI{43}{\kelvin} and \SI{85}{\kelvin} are highly sensitive to cold-working and heat treatment. 
Although high-temperature anneals reduce the effects of cold-working, even six-day anneals do not completely restore the sample's original magnetic properties.
Despite this sensitivity, measurements on known precipitates demonstrate that these transitions are intrinsic to the compound.  

This work emphasizes the need for subsequent investigations along multiple avenues.
Most immediately, future work might extend Hall experiments to lower temperatures and higher fields and examine specific heat under applied magnetic field.
Studies could also be undertaken to investigate the magnetic effects resulting from varying the composition of the Cantor alloy with the goal of controllably tuning the magnetic properties.
Lastly, the evidence presented in this work suggests that the distinctive processing dependence of the alloy is not due to the effects of known precipitates.
Lattice strain is a possible explanation, but a high-temperature anneal would be expected to eliminate strain effects.
That annealing does not restore the original state of the alloy suggests the presence of a mechanism that will require further research to elucidate.

\section{Acknowledgments} \label{ack}
Synthesis and characterization facilities at the University of Florida were developed under support from NSF-CAREER 1453752 (J.J.H.).  J.J.H.~benefited from enlightening conversations with D.L.~Maslov.
B.A.F.~and E.Z. acknowledge support from the College of Physical and Mathematical Sciences at Brigham Young University and also 
thank the scientific staff at the Centre for Molecular and Materials Science at TRIUMF for support during the muon spin relaxation experiment.
A portion of this work was performed at the National High Magnetic Field Laboratory, which is supported by the National Science Foundation Cooperative Agreement No. DMR-1644779 and the State of Florida. The authors acknowledge the staff of the Nano Research Facility at University of Florida for their assistance and guidance in acquiring SEM and XRD data.

The computational work was performed, in part, at the Center for Integrated Nanotechnologies, an Office of Science User Facility operated for the U.S.\ Department of Energy.
Sandia National Laboratories is a multimission laboratory managed and operated by National Technology \& Engineering Solutions of Sandia, LLC, a wholly owned subsidiary of Honeywell International Inc., for the U.S.\ Department of Energy’s National Nuclear Security Administration under contract DE-NA0003525.
This paper describes objective technical results and analysis. Any subjective views or opinions that might be expressed in the paper do not necessarily represent the views of the U.S.\ Department of Energy or the United States Government.
\bibliography{references,musr_bib}
\end{document}